# Convergence Behaviour of Bystanders: An Analysis of 2016 Munich Shooting Twitter Crisis Communication

**Deborah Bunker**
Business School
The University of Sydney
Sydney, Australia
Email: deborah.bunker@sydney.edu.au

**Milad Mirbabaie**
Department of Computer Science and Applied Cognitive Science
University of Duisburg-Essen
Duisburg, Germany
Email: milad.mirbabaie@uni-due.de

**Stefan Stieglitz**
Department of Computer Science and Applied Cognitive Science
University of Duisburg-Essen
Duisburg, Germany
Email: stefan.stieglitz@uni-due.de

## Abstract

While convergence behaviour archetypes can explain behaviour of individuals who actively converge on and participate in crises, less is known about individuals who converge on an event and choose to remain passive i.e. "bystanders". Bystanders are important, because of their proximity to an event and their function as an "eye-witness". To investigate the role of bystanders in crisis communications we analysed Twitter communication generated from the 2016 Munich Shooting event. Our findings revealed the *impassive* convergence behaviour archetype could influence an event as a passive and rational "eye-witness", by gathering and sharing information close to where the event is occurring.

# 1　Introduction

In the last decade, social media platforms have become platforms of mass communication and have revolutionised ways to communicate. For example, social networks, blogs, microblogs, wikis, and other types of social media concepts have had a tremendous impact on society (Chen et al. 2014; Li 2016) and business (Raisinghani 2012; Xu 2016).

For several reasons, social media platforms have become an important source of information for crisis management (Palen 2008; Pee 2012). Firstly, social media are used by individuals as a primary source of relevant crisis information at low cost (Sharif et al. 2013; Xu 2016). Secondly, social media have the ability to reach a large number of people in a short time and are considered as an important source of critical information by many people (Fischer et al. 2016, Eustace and Alam 2012; Freberg 2012). Using social media for sense-making and crisis management, however, also comes with some risks (Mirbabaie and Zapatka 2017; Stieglitz, Bunker, et al. 2017; Stieglitz, Mirbabaie, et al. 2017). For example, information can be ambiguous (Ahmed 2012) or rumours can be spread (Eustace and Alam 2012; Oh et al. 2013). On the one hand researchers and practitioners have made great progress in developing information systems that predict extreme events, such as earthquakes (Alvan and Azad 2011; Varnes 1989) and floods (Ruslan et al. 2016; Visconti et al. 2016). On the other hand, we are still struggling to manage human-driven extreme events, however, such as terrorist attacks and industrial accidents like oil spills. Existing information systems are mainly focussed on command and control responses to crises between a central command centre and local operational control during crisis event (Bunker et al. 2015) and so they can lack the necessary agility and flexibility to supply relevant information to frame local situational awareness for crisis and disaster management and response. Emergency management agencies (EMA) are also not yet able to effectively apply the methods and processes of social media analytics, to the high volumes of social media generated data, to generate relevant situational awareness for crisis situations in an efficient manner (van Gorp et al. 2015; Mukkamala and Beck 2016).

During an extreme event or a crisis, traditionally, communication between EMA and the general public was through mainstream media (newspapers, television, and radio) in a one-way direction. Nowadays social media platforms enable individuals and organisations to interact directly with each other (Blum et al. 2014). EMA can use social media in many ways (Subba and Bui 2017), for example. to spread information about an extreme event very quickly (Ehnis et al. 2014; van Gorp et al. 2015; Mirbabaie et al. 2014) or to source information from volunteers (Merrick and Duffy 2013; Stefanidis et al. 2013).

Extreme events and crises are highly complex but they are usually a "black box" when they first emerge, and where we also witness people communicating and interacting under uncertain conditions (Fischer et al. 2016; McKinney 2008). Yet, once a crisis is having a full impact, or when it has passed, people try to make sense of the situation exhibiting inquisitive and sense-making behaviours. Driven by the impact of the crisis on them or their close family members or friends, individuals search for and sometimes create information (Palen 2008; Pervin et al. 2014). Information and Communication Technologies (ICT), such as social media, allow individuals and organisations to communicate and make sense of interactions during crises considering different types of information, such as written texts, videos, images, and geographic information data (Schwarz 2012; Shahid and Elbanna 2015).

At the same time volunteers can support EMA and individuals who might be impacted by an extreme event by sharing geographic on-site information about the event, the situation and the conditions in the area (Horita et al. 2013; Mirbabaie et al. 2016). Recently, research has highlighted the power of Volunteered Geographic Information (VGI) which can be helpful in efficiently managing an extreme event (Fast and Rinner 2014). VGI is gaining a lot of attention, therefore, in both research and practice (Elwood et al. 2012; Merrick and Duffy 2013).

Crisis management is also complicated by human crisis (physical and virtual) convergence behaviour i.e. the spontaneous and mass movement of assets, people and resources towards an area impacted by a crisis. Convergence behaviour may have active or passive characteristics. For example, "disaster sightseers" may *physically converge* on an area where a crisis event is unfolding to see for themselves what is happening. While they have converged on the event they remain passive bystanders. In convergence behaviour theory, these individuals are referred to as "the curious" (Fritz & Mathewson 1957). In an online situation, a commonly available social media platform such as Twitter may facilitate the *virtual convergence* of individuals on a crisis and facilitate offers of assistance from them to affected and impacted individuals. These individuals converge on an event and are actively involved in a response to it. In convergence behaviour theory, these individuals are referred to as "the helpers". For instance, #ikwilhelpen was established during the Brussels terror attack in 2016 to enable "the helpers" to volunteer their assistance to those people impacted by the event (Perkins 2016).

While we understand the behaviour of individuals who actively converge on and participate in crises (both in the physical and virtual worlds) as well as the emerging roles in crisis communication networks which reinforce these behaviours, there is less knowledge about individuals who converge on an event and yet choose to remain a passive "bystander", i.e. people who are nearby an extreme event, not in any way, directly impacted by it, and who do not choose to participate in it other than to communicate about it. Bystanders are particularly important, because of their local proximity to an event and their function as "eye-witnesses", as well as their choice to remain a passive non-participant. It is important to better understand how these roles emerge and are reinforced by their social media communications. In developing this understanding we will know more about the impact and contributions of bystanders on the crisis communication process. Therefore, we seek to address the following research question: *How do crisis event "bystanders" such as the: anxious; curious; fans (or supporters); and mourners, utilise social media platforms to communicate during a crisis and does this have the potential to impact and influence an event?*

To answer the research question, we have analysed Twitter communication that was generated during the 2016 Munich Shooting crisis event. The remainder of the paper is structured as follows: Firstly, we describe the background and description of crisis communication and convergence behaviours in social media. Secondly, we describe our research design including the data collection, data preparation and data analysis. Thirdly, we summarise our findings and discuss them. Finally, we summarise our contribution and highlight aspects for further research.

## 2　Background

### 2.1　Crisis Communication in Social Media

One growing field of research for social media use during extreme events is 'communication and collaboration' (Arif et al. 2016; Oh et al. 2013; Olteanu et al. 2015). In this context two aspects are important: (1) social media can be utilised almost from everywhere and at any time and (2) social media allow everybody to spread information without verification (Kaplan and Haenlein 2010). In this regard, real-time dissemination of news is important, because it allows the user to stay informed and to spread news quickly (Raue et al. 2012; Zhao and Rosson 2009), which is especially relevant in uncertain situations. As a result, social media analytics can be applied to examine crisis communication management. Some examples can be found during the Red River Flood and the Oklahoma Fires in 2009 (Starbird and Palen 2010), the Queensland Flood 2011 (Bruns et al. 2012; Cheong and Cheong 2011; Shaw et al. 2013), the 2011 Tunisian Revolution (Kavanaugh et al. 2016) the Haiti Earthquake 2011 (Oh et al. 2010), the 2011 Norway Siege (Eriksson 2016), the 2011 Egypt Revolution and uprisings (Oh et al. 2015; Starbird and Palen 2012), Hurricane Sandy in 2012 (Gupta et al. 2013), the Boston Marathon Bombing 2013 (Cassa et al. 2013; Ehnis and Bunker 2013; Starbird et al. 2016), Typhoon Haiyan in the Philippines 2013 (Takahashi et al. 2015), and in context of the Sydney Siege 2014 (Archie 2016; Arif et al. 2016; Starbird et al. 2016). Starbird and Palen (2010) investigated tweets generated during the Red River Flood and Oklahoma Fires in 2009, which were co-occurring natural disasters in the USA. They demonstrated that tweets sourced by accounts of mainstream local media and service organisations were retweeted the most in both cases. Panagiotopoulos et al. (2016) published a study about Twitter usage during the heavy snowfalls in 2010 in the UK. The majority of the tweets generated dealt with local and crisis specific content like information about remaining resources of de-icing salt. Cassa et al. (2013) investigated the tweets occurring immediately after and around the Boston Marathon bombings in 2013. They revealed that the reports of bystanders, i.e. individuals who were in close proximity and witnesses dominated the Twitter communication in the first minutes of the event. Geo-located information provided through social media channels by citizens helped officials in localising the attacks (Cassa et al. 2013). Since most of the content on social media is generated by citizens and diffused through the network, a correct and complete understanding of a crisis cannot be guaranteed. This in turn highlights the need for identifying and analysing social media content continuously, to ensure successful sense-making by individuals, organisations like EMA (Mirbabaie and Zapatka 2017).

### 2.2　Crisis Convergence Behaviour

Crisis convergence behaviours have been recognised and researched for many years in the disaster and crisis management domain. Convergence behaviour is a well-known and well-researched phenomenon that occurs when a crisis or disaster happens. The mass movement of assets, people and resources towards the area that is impacted by the disaster, spontaneously occurs and individuals who converge on the disaster exhibit dominant behaviours. Fritz and Matthewson (1957) originally outlined 5 convergence behaviour "archetypes". These included the: *returnees; anxious; helpers; curious;* and *exploiters*. Kendra and Wachtendorf (2003) discovered two more archetypes from their analysis of the 9/11 terrorist attacks in 2001 i.e. *fans* or *supporters*; and *mourners*. Subba and Bui (2010) then added one more archetype from their study of physical and virtual (online) convergence behaviours i.e. the *detectives* while Bunker and Sleigh (2016) have recently proposed an additional convergence behavioural archetype of the *manipulator* which has evolved as a result of the development and adoption of social media platforms and their capacity to enable narcissistic and anti-social behaviour during a crisis event (see Table 1).

**Table 1.** Convergence Behaviour Archetypes – Bunker and Sleigh (2016)
*- originally adapted from Subba and Bui, 2010*

| Authors | Convergence Behaviour Archetype | Characteristics |
|---|---|---|
| Fritz and Matthewson, 1957 | The returnees | Strong sense of legitimacy to enter a disaster area e.g. evacuated residents, friends and family of residents, property owners - many and strong motivations to return. |
| Fritz and Matthewson, 1957 | The anxious | Fall into 2 categories - anxious close associates of those directly impacted by the disaster, generally anxious about those affected by the disaster. Sub-categorized as information *seekers* and *responders*. |
| Fritz and Matthewson, 1957 | The helpers | Volunteer to help disaster victims and fall into sub-categories of formal (PSA) and informal (everyone else). |
| Fritz and Matthewson, 1957 | The curious | Minimal personal concerns i.e. "sightseeing". |
| Fritz and Matthewson, 1957 | The exploiters | Looking for personal gain, detachment from or non-sympathetic identification with the victims. Manifesting in scamming, looting, stealing, giving misleading information etc. |
| Kendra and Wachtendorf, 2003 | The fans or supporters | Encourage or express gratitude to rescuers. |
| Kendra and Wachtendorf, 2003 | The mourners | Memorialize and mourn the dead. |
| Subba and Bui, 2010 | The detectives | Official and unofficial intelligence gatherers who watch over activities and take appropriate action. |
| Bunker & Sleigh 2016 | The manipulators | Looking to promote self and project personal characteristics of power, intelligence, physical attractiveness, sense of entitlement and uniqueness. Manifests in attention seeking behaviour and creating or seeking roles of perceived importance in the management of the disaster. |

When we closely examine these convergence behaviours we see that they fall into two different role types of those who have: 1) *active crisis involvement* i.e. returnees; helpers, exploiters; detectives or manipulators; or (2) *passive crisis bystander,* i.e. anxious; curious; fans (or supporters); and mourners status. Subba and Bui (2010) also argue that all convergence behaviours have physical and virtual interaction "properties" which include: *Local V Global* e.g. a local event may have a global impact and vice versa; *Complementarity V Substitutability* e.g. roles such as first responders may be enacted online but may not necessarily substitute for first responders at the scene; *Formality V Informality* e.g. first responder agencies have a formal response role while communities and individuals may have a less formal response role; *Legitimacy v Illegality* e.g. responses to a crisis event that are desirable, proper and appropriate versus those that are illegal or morally wrong; *Planning V Spontaneity* e.g. planned and formal responses versus ad hoc and emergent responses such as agency trained volunteers or individual "spontaneous volunteers"; and *Centralized V Decentralized* e.g. responses at the crisis site versus collaboration and co-operation via social media platforms. Subba and Bui (2010) highlight that Fritz and Matthewson (1957) explain that an "initial attack on the problems of convergence requires the development of a systematic policy and programs for handling information and communications in disasters" (Subba & Bui, 2010 - page 9) but because social media platforms produce information and communication in a haphazard, organic and disorganised manner they often produce emergent, persistent, undesirable and unwanted convergence behaviour.

# 3 Research Design

## 3.1 Case Description

The tragedy of the Munich shooting occurred on 22 July 2016 at 3:52pm UTC west of the Olympia shopping mall in Munich, Germany. A single 18-year-old German man caused the shooting. Ten people were killed (including the perpetrator) and 36 civilians were injured. We studied the Munich shooting as an exemplar case due to its geographic specificity i.e. not spread over a wide area, as well as its direct impact on a few individuals. In this way we could effectively classify tweets that were generated by individuals into *active crisis involvement* and *passive crisis bystanders*. This case was also selected for analysis because it was accompanied by an intense online discussion on social media platforms including many assumptions and rumours about the number of attackers, the exact location(s), and the number and identity of the victims. After the event, the police confirmed that a single mentally ill person perpetrated the shooting but it was firstly assumed by social media participants, that the shooting was a terrorist attack. Social media, especially Twitter and Facebook, were highly utilised as communication tools to make sense of the crisis situation during and after this event.

## 3.2 Data Collection

For the purposes of our study we focussed on the microblogging platform Twitter, as it was an important channel for communication and cooperation for both crisis management agencies and the public. Twitter was selected, because it offers an API for collecting data, i.e. tweets and retweets, but also meta-data, such as information about the author of a tweet or the number of followers (Bruns and Liang 2012). The platform is also known for its fast and spontaneous information diffusion, which is reflected in its 313[1] million monthly active users worldwide. Local media correspondents also tend to use Twitter to publish breaking news to the location of crises and emergency incidents (Oh et al. 2012).

We also used a self-developed Java tool, for data tracking, which connects to the Search API[2], offered by Twitter. Tweets can only be collected by defining keywords or by tracking specific accounts. For this case study, we chose both opportunities and defined keywords and tracked the official account of the police in Munich.

We focused on tweets, which were declared (by the platform Twitter itself) as German. This selection was justified by the motivation of only analysing tweets, which were produced in the area of the shooting in Germany. By following the ranking algorithm of Twitter, suggesting which keywords and hashtags are trending and by observing the communication for several hours, we were able to identify the following keywords: "*münchen*", "*prayformunich*", "*munich*", "*oez*". We also collected all tweets, which were published by the police in Munich: "*@polizeimuenchen*". The Search API offered the opportunity to track data retroactively (up to one week in the past, depending on the volume). As we started to track the data directly after the first news about the shooting, we were able to collect those tweets, which were published right before and at the moment of the shooting. The data that was collected was written continuously in a MySQL database, from where we later conducted the analyses.

The data collection timeframe was from 22 July 2016 0am UTC until 25 July 2016 0am UTC, i.e. before the shooting event and after the crisis was concluded. Overall, we collected 672,871 tweets related to this event.

## 3.3 Data Preparation and Analysis

In order to identify the passive bystanders, i.e. people who were located in the area of the shooting, but did not have an active role in the event i.e. *anxious; curious; fans (or supporters); and mourners,* we extracted only tweets with GPS coordinates. As a result, we constructed a list with 1,651 tweets including GPS coordinates such as longitude and latitude, provided by the Search API of Twitter. For this case study, we selected only those tweets, which were written from a maximum distance of 10 km to the exact location of the event (48°11′0″N 11°32′1″E). Due to the geographic specificity of the incident we determined that the 10km radius would be far out enough to identify bystanders, but not too far away from the epicentre of the incident.

---

[1] https://about.twitter.com/company, last access: 2017-18-01

[2] https://dev.twitter.com/rest/public/search, last access: 2017-18-01

By conducting our research as follows, we are able to identify the bystanders, which enabled us to answer our research question: *How do crisis event "bystanders" such as the anxious, curious, fans (or supporters) and mourners utilise social media platforms to support these convergence behaviours?*

**Firstly**, the tweets were filtered by time, the use of GPS-coordinates, as well as certain hashtags and then exported into a CSV file. After that, we wrote a python script that calculated the distance between the location of each tweet and the location of the shooting using the extern library „*geopy*[3]". Every tweet that was not written within a radius of 10,000 metres was removed from our dataset.

**Secondly**, by manual coding of profile information and posted content, we categorised the authors of these tweets into the following predefined convergence behaviour archetypes of the: 1) returnees (active); anxious (passive); helpers (active); curious (passive); exploiters (active); fans or supporters (passive); mourners (passive); detectives (active); and manipulators (active).

**Thirdly**, we searched for new archetypes, which did not fit with the previously defined archetypes.

## 4 Findings

Our overall analysis of this data highlighted that most of the individuals, who participated in the Twitter crisis communication during this event were private individuals, however, a few professional news producers and some public figures also participated in these communications. Profit driven organisations were also visible in our overall analysis, as they used one of the keywords we used for tracking the data, but most of this Twitter content was not crisis relevant. One explanation could be that the usage of trending keywords, such as *Munich* (in our case) were utilised by organisations to get the attention of the social network regarding a brand/and or company. In one case a profit driven organisation published crisis related content, which contained a picture and an appeal to fight against terrorism. In all social network analysis and timeframes of the crisis, it is clear that most of the bystanders were located in the Munich inner city area and were not at the centre of the shooting. A few private individuals, however, could be detected in the general area of the shooting.

**Table 2.** Frequencies of Archetypes according to their distance to the incident

| Archetype | Within 10 km area | | More than 10 km | | Total | |
|---|---|---|---|---|---|---|
| | Number | Percentage | Number | Percentage | Number | Percentage |
| Bots | 0 | 0 | 268 | 100.00 | 268 | 17.57 |
| The Anxious (passive) | 15 | 12.50 | 35 | 87.50 | 50 | 3.28 |
| The Curious (passive) | 2 | 7.80 | 23 | 92.20 | 25 | 1.64 |
| The Exploiters (active) | 1 | 2.50 | 39 | 97.50 | 40 | 2.62 |
| The Furious (passive) | 1 | 2.22 | 44 | 97.78 | 45 | 2.95 |
| The Helpers active (informal) | 5 | 38.46 | 8 | 61.54 | 13 | 0.85 |
| The Impassive (passive) | 312 | 72.39 | 119 | 27.61 | 431 | 28.26 |
| The Informers (active) | 19 | 5.05 | 357 | 94.95 | 376 | 24.66 |
| The Manipulators (active) | 19 | 38.78 | 30 | 61.22 | 49 | 3.21 |
| The Mourners (passive) | 33 | 31.13 | 73 | 68.87 | 106 | 6.95 |
| The Promoters (passive) | 35 | 72.92 | 13 | 27.08 | 48 | 3.15 |
| The Returnees (active) | 1 | 50.00 | 1 | 50.00 | 2 | 0.13 |
| The Supporters/Fans (passive) | 21 | 29.17 | 51 | 70.83 | 72 | 4.72 |

---

[3] https://github.com/geopy/geopy, last access: 2017-18-01

After performing this overall analysis of the Twitter dataset we firstly, analysed the frequency of tweets by the two different convergence behaviour archetype categories i.e. *active crisis involvement* and *passive crisis bystander* as well as by distance from the crisis event i.e. within a 10km and outside of a 10km radius (see Table 2).

We found that communication increased markedly outside of the 10 km radius across all active and passive convergence behaviour archetypes (other than *impassive - passive* where the number of tweets were of greater volume inside the 10km radius*)* and that there were in excess of more than 100 tweets in total in the passive bystander archetypes of *impassive* (431) and *mourner* (106). There were also in excess of 100 tweets by the *informer (active)* archetype (376). It is also worth noting that the *impassive (passive)* and *informer (active)* archetypes are newly created archetypes from our analysis of data in this study. We note that the *informer (active)* category, however, may be a sub-set of the *detective* convergence behavioural archetype. The convergence behaviour archetypes highlighted in Table 2, also align with our analysis illustrated in Fig.1, which shows the overall percentage of tweets by convergence behaviour archetypes.

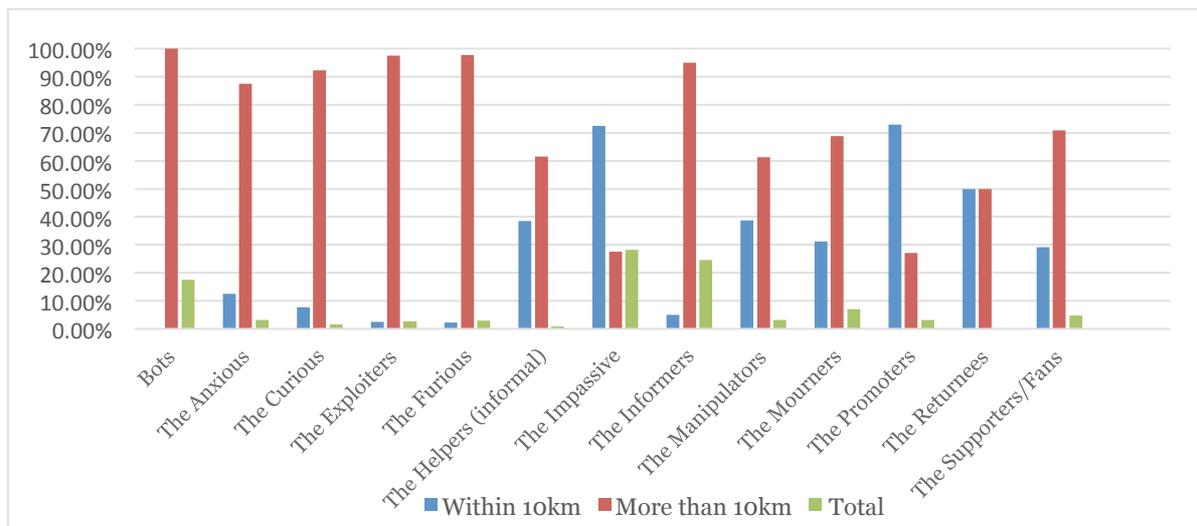

**Fig 1.** Frequencies of Archetypes according to their distance to the incident

We then analysed the percentage of tweets highlighting information content themes i.e. personal, location, trend, solicitousness, crisis, help/shelter, other news, media, opinion, advertisement and other, by active crisis management and passive crisis bystander archetypes both within a 10km radius (Fig. 2) and outside of a 10km radius (Fig. 3).

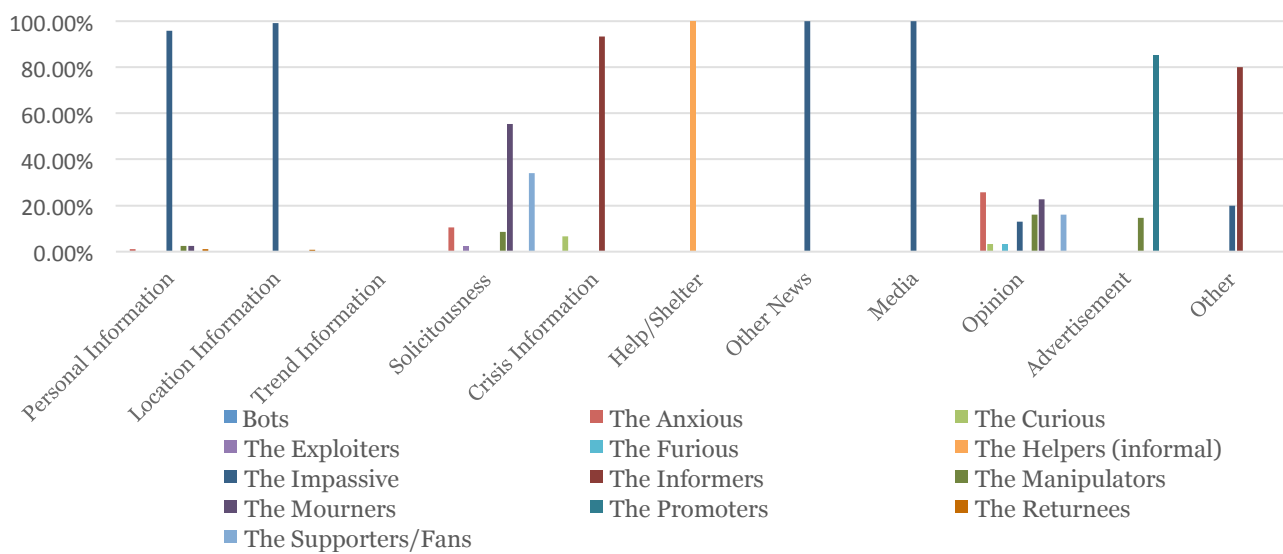

**Fig. 2.** Archetypes and Types of Information inside of a radius of 10 km

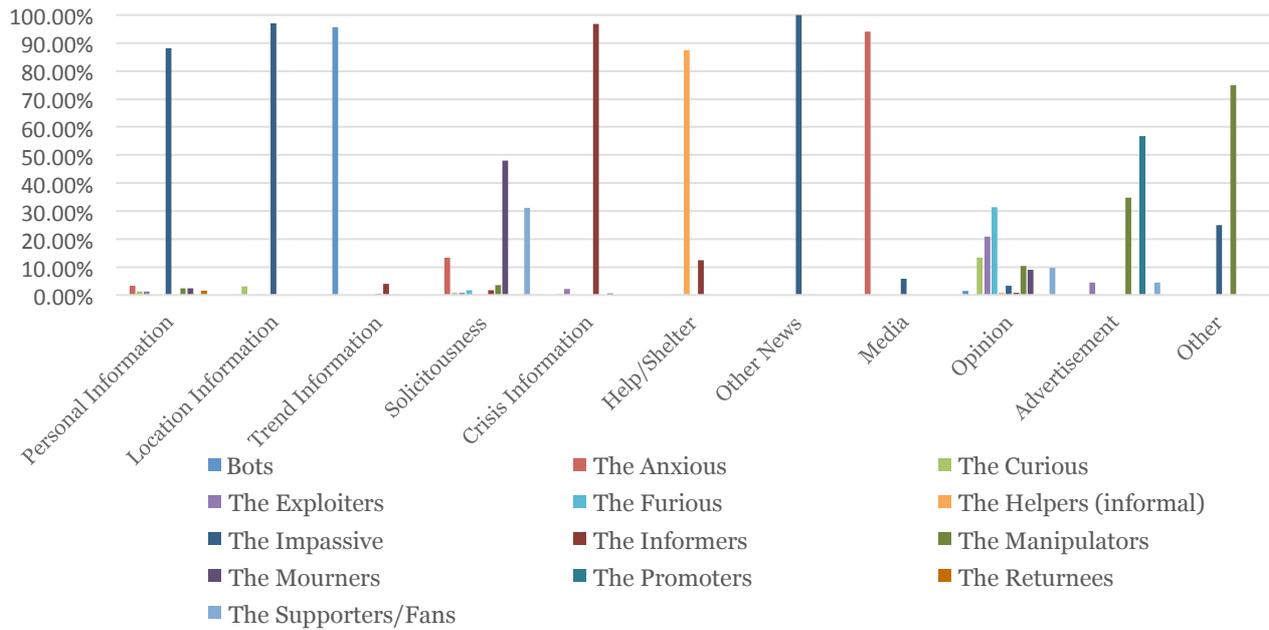

**Fig. 3.** Archetypes and Types of Information outside of a radius of 10 km

Within a 10 km radius *the impassive (passive)* were focused on personal, location, other news and media themes. The *informers (active)* were focussed on crisis and other information, while *helpers (active)* focussed on help/shelter and *promoters (passive)* focussed advertisement information.

In our analysis of the percentage of tweets outside of a 10km radius we found that *the impassive (passive)* were focused on personal, location and other news themes. The *informers (active)* were focussed on crisis information, while *helpers (active)* focussed on help/shelter; the *anxious (passive)* on media information and *promoters (passive)* focussed advertisement information. The *manipulators (active)* were focussed on other information.

## 5　DISCUSSION

So how do crisis event "bystanders" such as the: anxious; curious; fans (or supporters); and mourners, utilise social media platforms to communicate during a crisis and does this have the potential to impact and influence an event?

Firstly, we were able to observe the emergence of 5 new convergence behaviour archetypes that seem to have been enabled by the use of social media for crisis communications i.e. Furious (passive), Impassive (passive), Promoters (passive) as well as the Exploiters (active) and Informers (active).

Overall Twitter communication did increase the further away from the event it occurred whether the convergence behaviour archetype was one of *active* crisis involvement or *passive* crisis bystander. This was consistent for all archetypes other than the *impassive (passive)* where the number of tweets was of greater volume inside the 10km radius and their content was focussed on personal, location, and other news and media themes. *Mourners (passive)* showed a large percentage increase in tweets the further away from the event that they were and these tweets were focussed on content that addressed solicitousness.

Outside of a 10km radius we found that while *the impassive (passive)* archetype tweet volume decreased these individuals were still focused on personal, location and other news themes. The media theme dropped out of immediate concern the further away from the incident that these individuals were located. The *anxious (passive)* focussed on media information and *promoters (passive)* focussed advertisement information the further away from the event that they were.

In general we observed that the percentage of tweet content changed very little from within and outside of the 10km radius by archetype category.

The surprising finding from the analysis, however, was the emergence of the impassive (passive) convergence archetype, their proximity to the event, the volume of tweets that they generated inside of the 10km radius as well as the content of these tweets. All of these factors combined i.e. proximity to the event, volume and content (of communications), highlight that this bystander archetype may have the ability to significantly influence an event by bringing together important information in a non-emotional/rational way, on personal, location, other news and media themes close to the emerging crisis. These individuals have the potential to provide a valuable function in emergency management communications if we can better understand this convergence behaviour archetype utilises social media platforms for crisis communications.

# 6　CONCLUSION AND OUTLOOK

## 6.1　Conclusion

In our paper, we have conducted a case study to identify bystanders in social media crisis communication and their "eye-witness" contribution to the crisis management process. For this purpose, we focused on the 2016 Munich Shooting by collecting the communication on the microblogging platform Twitter. By using several analysis methods, we were able to locate bystanders via GPS data and highlight their contribution to crisis communications during an event. Our paper is the first step towards the creation of a common understanding of bystander crisis communication and its potential influence and impact on a crisis event.

## 6.2　Contributions

The contribution of this paper lies in the analysis of the bystander communication in both close and not so close proximity to a crisis event. We enriched the convergence behaviour archetype literature by revealing the emergence of new archetypes through our analysis of the Twitter crisis communication dataset. Our findings could be useful for organisations, such as EMA to make sense of the bystander communication and how this might be utilised to influence the outcomes and management of such an event.

## 6.3　Limitations and Further Research

We are aware of the limitations of this research. First, our sample size is small. This was caused by the nature of the platform (Twitter) and also that individual users do not automatically turn on their GPS module on their mobile phones due to privacy concerns. Second, Twitter is our source for analysing the bystander communication and possibly other social media platforms might be also be relevant to consider in addition in order to gain supplementary, more holistic and meaningful insights.

Further research in this area will therefore be extended to: 1) other types of crises; and 2) other social media platforms, such as Facebook. We will also consider other possibilities of identifying bystanders, such as personally nominated characteristics. Another approach would be to use Natural Language Processing to analyse the exact location of individuals from the social media communications and content they produce. Besides social media platforms, other crisis crowdsourcing systems can be considered, such as Ushahidi, to compare the activities of convergence archetypes with active crisis involvement with those of passive crisis bystanders.

# 7　References